\documentclass[12pt]{article}
\usepackage{graphicx}
\usepackage{amsmath}

\newtheorem{theorem}{Theorem}[section]

\newtheorem{lemma}[theorem]{Lemma}

\newenvironment{proof}[1][Proof]{\textsc{#1.} }{\ \rule{0.5em}{0.5em}}
\numberwithin{equation}{section}
\input{tcilatex}

\begin{document}

\title{Future complete $S^{1}$ symmetric solutions of the Einstein Maxwell Higgs
system.}
\author{Yvonne Choquet-Bruhat}
\maketitle

\section{Introduction.}

We have obtained in previous works [CB-M1], [CB-M2], [CB2] a proof of the
non linear stability (i.e. future completeness for small initial data) of
Einsteinian vacuum spacetimes with compact space sections and a $S^{1}$
spatial isometry group, in the case where the space is an $S^{1}$ bundle
over a surface $\Sigma $ of genus greater than one. We intend in this paper
to extend this result to the Einstein Maxwell Higgs system. Our previous
result is proved by using the fact that the 4 dimensional vacuum Einstein
equations with a one parameter isometry group are essentially (i.e. up to a
harmonic one form on $\Sigma $ which we chose to be zero) equivalent to an
Einstein - wave map system on the quotient 3 dimensional manifold, the
Einstein equations being equivalent to a time dependent elliptic system on $%
\Sigma $ coupled in the case of genus($\Sigma )>1$ with ordinary
differential equations governing the evolution of conformal classes of the
time dependent metric of $\Sigma .$ The appearance of a harmonic map
generalizing the Ernst equation in the stationary Einstein Maxwell system
has been found by Kinnersley and Mazur (see [Ma]) who studied the associated
invariance group. The reduction of the Einstein Maxwell system, with a
spatial $S^{1}$ isometry group to a coupled Einstein wave map system on
three dimensional space has been proved by Moncrief [Mo1] in the case $%
\Sigma =S^{2},$ and for a class of Einstein Maxwell Higgs equations in the
case $\Sigma =T^{2}$. Though the notations and formalism used in these
reductions were different from the one with which we obtained our global
existence proof, it was likely, that the proof of future completeness
obtained for the vacuum unpolarized Einstein equations in [CB2] with $\Sigma 
$ of genus greater than one will extend to the Einstein Maxwell Higgs system.

It is known since the works of Kaluza and Klein that one can recover the
Einstein Maxwell equations by considering a Lorentzian metric $^{(5)}g$ on a
five dimensional manifold $V_{5}$ with a spacelike 1 - parameter isometry
group $G_{1},$ writing the vacuum Einstein equations for $(V_{5},^{(5)}g)$
and discarding a wave type equation for a scalar field taken to be constant.
If one considers the full equations\footnote{{\footnotesize See for instance
[L]. \ }} one finds them equivalent to the Einstein Maxwell equations with
an additional scalar field $\phi $ which interacts with both gravitation and
electromagnetism. If the gravitational and electromagnetic fields as well as
the additional scalar field are themselves invariant under another spatial
isometry group, then the 5 dimensional metric $^{(5)}g$ is invariant under a
2 - parameter abelian group $G_{2}$ endowing $V_{5}$ with a $G_{2}$ fiber
bundle structure with base a manifold $V_{3}.$ In this article we consider
more generally a $d$ dimensional manifold $V_{d}$ with a Lorentzian metric $%
^{(d)}g$ which admits an $m$ dimensional isometry group which endows it with
a fiber bundle structure with base an $n=d-m$ dimensional manifold $V_{n}$.
We write the general Kaluza Klein formulas expressing the Ricci tensor of $%
^{(d)}g$ in terms of geometric elements defined on $V_{n},$ in the case of
an abelian isometry group. We show that in the case $n=3$ the vacuum
Einstein equations in dimension $d$, $Ricci(^{(d)}g)=0,$ are (modulo the
choice zero for $d-m$ harmonic one forms) equivalent to a coupled Einstein
wave map system on $V_{3}$. It gives in particular by taking $d=5$ the
Einstein wave map system which is equivalent to the $4$ dimensional Einstein
equations with Maxwell - Higgs sources and $S^{1}$ isometry group. We
finally sketch the extension to that system of the proof used for the non
linear stability theorem [CB2] of the vacuum unpolarized $4$ dimensional
Einstein equations with $S^{1}$ isometry group.

\section{General Kaluza Klein equations.}

If the Lorentzian metric $^{(d)}g$ on the $d$ manifold $V_{d}$ admits a $d-n$
dimensional isometry group $G_{d-n}$ endowing $V_{d}$ with a fiber bundle
structure with base the $n$ manifold $V_{n}$ this metric $^{(d)}g$ can be
written as follows in a local trivialization of the bundle over an open set
of $V_{n}$: 
\begin{equation}
^{(d)}g\equiv \underline{g}+\xi _{mn}(\theta ^{m}+a^{m})(\theta ^{n}+a^{n}),
\end{equation}
where \underline{$g$} is a Lorentzian metric on $V_{n},$ $\theta ^{m}$ a
basis of invariant 1 forms on $G_{d-n},$ $a\equiv (a^{m})$ the representant
on $V_{n}$ of a $G_{d-n}$ connection on $V_{d},$ i.e. a 1 form with values
in $\mathcal{G}$, the Lie algebra of $G_{d-n},$ we denote by $F\equiv
(F^{m}) $ its curvature 2 form. In local coordinates on $V_{n}$ we have:

\begin{equation}
\underline{g}\equiv \underline{g}_{\alpha \beta }dx^{\alpha }dx^{\beta },%
\text{ \ }a^{m}\equiv a_{\alpha }^{m}dx^{\alpha }.
\end{equation}
We suppose that, as in the example given above, $G_{d-n}$ is an \textbf{%
abelian group} then: 
\begin{equation}
F=da,\text{ \ \ i.e. \ \ }F_{\alpha \beta }^{m}=\partial _{\alpha }a_{\beta
}^{m}-\partial _{\beta }a_{\alpha }^{m}
\end{equation}
and the gauge covariant derivatives on sections of vector bundles,
associated with the principal fiber bundle with group $G_{d-n}$ over $(V_{n},%
\underline{g}),$ coincide with the usual covariant derivatives in the metric 
$\underline{g}$

We recall the Kaluza-Klein form of the Ricci tensor of $^{(d)}g$ (see CB-DM
II , V 13) and use it to write the vacuum Einstein equations on $%
(V_{d},^{(d)}g)$.

We set: 
\begin{equation}
(\det \xi )^{\frac{1}{2}}=:e^{X},\text{ \ \ \ hence \ \ }\xi ^{mn}\partial
_{\alpha }\xi _{mn}=2\partial _{\alpha }X,
\end{equation}
then the general Kaluza Klein formulas give the following system, all
quantities being fields on $V_{n}$ with greek indices raised with $%
\underline{g}$ and latin indices $m,n,...$ with $\xi :$ 
\begin{equation}
^{(d)}R_{\alpha \beta }\equiv \underline{R}_{\alpha \beta }-\underline{%
\nabla }_{\alpha }\partial _{\beta }X+\frac{1}{4}\partial _{\alpha }\xi
_{mn}\partial _{\beta }\xi ^{mn}+\frac{1}{2}F_{m,\alpha }{}^{\lambda }%
\underline{F}_{\beta \lambda }^{m}=0
\end{equation}
\begin{equation}
^{(d)}R_{\alpha m}\equiv -\frac{e^{-X}}{2}\underline{\nabla }_{\lambda }[%
\underline{F}_{m\alpha }{}^{\lambda }e^{X}]=0
\end{equation}
\begin{equation}
^{(d)}R_{mn}\equiv -\frac{1}{2}[\underline{\nabla }^{\alpha }\partial
_{\alpha }\xi _{mn}+\underline{\partial }^{\alpha }X\partial _{\alpha }\xi
_{mn}]+\frac{1}{2}\xi ^{pq}\partial _{\alpha }\xi _{mp}\underline{\partial }%
^{\alpha }\xi _{nq}-\frac{1}{4}F_{m,\alpha \beta }\underline{F}_{n}^{\alpha
\beta }=0.
\end{equation}

We see that the equations $^{(d)}R_{mn}=0$ are a quasidiagonal semilinear
sytem of wave equations for the field $\xi \equiv ($ $\xi _{mn}),$ they are
first order in the derivatives of the other unknowns $\underline{g}$ and $a.$

The equations $^{(d)}R_{\alpha m}=0$ look like Maxwell type equations for $%
F, $ with a weight $e^{X}.$

The equations $^{(d)}R_{\alpha \beta }=0$ look like Einstein equations with
source the Maxwell and scalar fields, except for the appearance of second
derivatives of $X.$ The combination of these equations with the system $%
^{(d)}R_{mn}=0$ takes the form of usual Einstein equations with sources by
introduction of a metric $\tilde{g}$ conformal to $\underline{g}.$ We will
perform the explicit computation in the case of interest to us, $n=3,$ which
permit the introduction of twist potentials to solve the Maxwell type
equations.

\section{Case n=3.}

\subsection{Twist potentials, definition.}

The equations 2.6 are equivalent to 
\begin{equation}
dE_{m}=0\text{ \ \ \ with \ \ \ }E_{m}=:\underline{\ast }[F_{m}e^{X}]
\end{equation}
with \underline{$\ast $} the adjoint operator on forms in the metric $%
\underline{g}.$ When $n=3$ the adjoint $E_{m}$\ of the 2 form $F_{m}$ is a 1
form. The general solution of 2.6 is then: 
\begin{equation}
E_{m}=d\omega _{m}+H_{m}
\end{equation}
with $\omega _{m}$ an arbitrary scalar function on $V_{3},$ called a twist
potential, and $H_{m}$ a representant of a 1 cohomology class. We consider
the class of solutions such that $H_{m}=0.$

\subsection{Associated tensors.}

The inversion of the defining equations for $E_{m}$ gives 
\begin{equation}
F_{m}=e^{-X}\underline{\ast }E_{m}=e^{-X}\underline{\ast }(d\omega _{m})
\end{equation}
that is with \underline{$\eta $} the volume form of \underline{$g$}: 
\begin{equation}
F_{m,\alpha \beta }=\underline{\eta }_{\alpha \beta \lambda }e^{-X}%
\underline{g}^{\lambda \mu }\partial _{\lambda }\omega _{m}\text{ \ \ \ \
hence \ \ \ }F_{\alpha \beta }^{m}=\underline{\eta }_{\alpha \beta \lambda
}e^{-X}\underline{g}^{\lambda \mu }\xi ^{mn}\partial _{\lambda }\omega _{n}.
\end{equation}

\begin{lemma}
The following identities hold (underlining means that greek indices are
raised with \underline{$g$}$^{\alpha \beta }):$ 
\begin{equation}
\underline{F}_{m,\alpha }{}^{\lambda }F_{n,\beta \lambda }\equiv e^{-2X}[%
\underline{g}_{\alpha \beta }\partial _{\lambda }\omega _{m}\underline{%
\partial }^{\lambda }\omega _{n}-\partial _{\alpha }\omega _{m}\partial
_{\beta }\omega _{n}]
\end{equation}
\begin{equation}
\underline{F}_{m}{}^{\lambda \mu }F_{\lambda \mu }^{m}\equiv 2e^{-2X}\xi
^{mn}\partial _{\lambda }\omega _{m}\underline{\partial }^{\lambda }\omega
_{n}.
\end{equation}
\end{lemma}

\begin{proof}
The first identity is obtained for instance by a computation in orthonormal
frame, the second results by contraction on the 3 manifold $V_{3}.$
\end{proof}

\subsection{Equations.}

The 2 forms $F^{m}$ being the differentials of the 1 forms $a^{m}$ are
necessarily closed, hence the functions $\omega _{m}$ must satisfy the
equations 
\begin{equation}
d[\underline{\ast }e^{-X}\xi ^{mn}d\omega _{m}]=0
\end{equation}
equivalently the functions $\omega _{m}$ must satisfy the following
semilinear wave equations on $(V_{3},\underline{g}):$ 
\begin{equation}
\underline{\nabla }^{\alpha }[e^{-X}\xi ^{mn}\partial _{\alpha }\omega
_{m}]=0,\text{ }.
\end{equation}

\section{Conformal metric.}

We introduce on $V_{3}$ the conformal metric 
\begin{equation}
\tilde{g}_{\alpha \beta }=:e^{2X}\underline{g}_{\alpha \beta },\text{ \
hence \ \ \ \underline{$g$}}^{\alpha \beta }=e^{2X}\tilde{g}^{\alpha \beta }.
\end{equation}

A simple calculation shows that the divergence of covariant vectors $v$ are
linked by the identity 
\begin{equation}
\underline{\nabla }_{\alpha }\underline{v}^{\alpha }\equiv e^{2X}(\tilde{%
\nabla}_{\alpha }\tilde{v}^{\alpha }-\tilde{v}^{\alpha }\partial _{\alpha
}X),\text{ \ \ }\underline{v}^{\alpha }=e^{2X}\tilde{v}^{\alpha }.
\end{equation}
In particular the wave operators in the metrics $\underline{g}$ and $\tilde{g%
}$ are linked by the identity 
\begin{equation}
\underline{\nabla }_{\alpha }\underline{\partial }^{\alpha }\equiv e^{2X}%
\tilde{\nabla}_{\alpha }\tilde{\partial}^{\alpha }-\underline{\partial }%
^{\alpha }X\partial _{\alpha }.
\end{equation}

\subsection{Scalar equations.}

We deduce from 2.7 that 
\begin{equation}
^{(d)}R_{mn}\equiv -\frac{e^{2X}}{2}[\tilde{\nabla}^{\alpha }\partial
_{\alpha }\xi _{mn}+\xi ^{pq}\partial _{\alpha }\xi _{mp}\tilde{\partial}%
^{\alpha }\xi _{nq}]-\frac{1}{4}F_{m,\alpha \beta }\underline{F}_{n}^{\alpha
\beta }]=0.
\end{equation}
where $F_{m,\alpha \beta }\underline{F}_{n}^{\alpha \beta }$\ is given by
3.5.

These equations and $X=\log (\det \xi )^{1/2}$ imply by a simple computation
that: 
\begin{equation}
\xi ^{mn(d)}R_{mn}\equiv -e^{2X}\tilde{\nabla}_{\lambda }\tilde{\partial}%
^{\lambda }X+\frac{1}{4}F_{m,\alpha \beta }\underline{F}^{m,\alpha \beta }=0,
\end{equation}
that is, using 3.6: 
\begin{equation}
\tilde{\nabla}_{\lambda }\tilde{\partial}^{\lambda }X-\frac{1}{2}e^{-2X}\xi
^{mn}\partial _{\lambda }\omega _{m}\tilde{\partial}^{\lambda }\omega _{n}=0.
\end{equation}
For simplicity of proofs we will sometimes introduce the scalar function $%
X=\log (\det \xi )^{1/2}$ as an auxiliary unknown. The equations 4.4 and 4.6
will imply that 
\begin{equation}
X=\frac{1}{2}logdet\xi
\end{equation}
if this property and its first time derivative are satisfied initially.

\subsection{Twist potentials equations.}

We have the identities: 
\begin{equation}
\underline{\nabla }^{\alpha }[e^{-X}\xi ^{mn}\partial _{\alpha }\omega
_{m}]\equiv e^{2X}\{\tilde{\nabla}^{\alpha }[e^{-X}\xi ^{mn}\partial
_{\alpha }\omega _{m}]-\tilde{\partial}^{\alpha }X(e^{-X}\xi ^{mn}\partial
_{\alpha }\omega _{m})]\}
\end{equation}
\begin{equation}
\equiv e^{3X}\tilde{\nabla}^{\alpha }[e^{-2X}\xi ^{mn}\partial _{\alpha
}\omega _{m}]\equiv e^{X}\xi ^{mn}\tilde{\nabla}^{\alpha }[\partial _{\alpha
}\omega _{m})]+e^{X}[\tilde{\partial}^{\alpha }\xi ^{mn}-2\xi ^{mn}\tilde{%
\partial}^{\alpha }X]\partial _{\alpha }\omega _{m}
\end{equation}
The equations 3.8 for the twist potentials are therefore: 
\begin{equation}
\tilde{\nabla}^{\alpha }\partial _{\alpha }\omega _{p}+\xi _{np}[\tilde{%
\partial}^{\alpha }\xi ^{mn}-2\xi ^{mn}\tilde{\partial}^{\alpha }X]\partial
_{\alpha }\omega _{m}=0
\end{equation}
that is 
\begin{equation}
\tilde{\nabla}^{\alpha }\partial _{\alpha }\omega _{p}-\xi ^{mn}(\tilde{%
\partial}^{\alpha }\xi _{np}+2\tilde{\partial}^{\alpha }X)\partial _{\alpha
}\omega _{m}=0
\end{equation}

\subsection{Einstein equations on $V_{3}.$}

On the $3$ dimensional manifold $V_{3}$ it holds that (see CB-DM I p351): 
\begin{equation}
\underline{R}_{\alpha \beta }\equiv \tilde{R}_{\alpha \beta }+\underline{%
\nabla }_{\alpha }\partial _{\beta }X-\partial _{\alpha }X\partial _{\beta
}X+\underline{g}_{\alpha \beta }(\underline{\nabla }^{\lambda }\partial
_{\lambda }X+\underline{\partial }^{\lambda }X\partial _{\lambda }X)
\end{equation}
We deduce from 2.5 and 4.12 that the term \underline{$\nabla $}$_{\alpha
}\partial _{\beta }X$ disappears from $^{(d)}R_{\alpha \beta }.$ We find
that: 
\begin{equation}
^{(d)}R_{\alpha \beta }\equiv \tilde{R}_{\alpha \beta }+\frac{1}{4}\partial
_{\alpha }\xi _{mn}\partial _{\beta }\xi ^{mn}-\frac{1}{2}\underline{F}%
_{m,\alpha }{}^{\lambda }F_{\beta \lambda }^{m}-\partial _{\alpha }X\partial
_{\beta }X+\tilde{g}_{\alpha \beta }\tilde{\nabla}^{\lambda }\partial
_{\lambda }X
\end{equation}
and we deduce from 4.5 that the combination $^{(d)}R_{\alpha \beta }+e^{-2X}%
\tilde{g}_{\alpha \beta }\xi ^{mn}{}^{(d)}R_{mn}=0$ reduces to the following
Einstein equations on $(V_{3},\tilde{g}):$ 
\begin{equation}
\tilde{R}_{\alpha \beta }=\rho _{\alpha \beta }
\end{equation}
with, using 3.5: 
\begin{equation}
\rho _{\alpha \beta }\equiv -\frac{1}{4}\partial _{\alpha }\xi _{mn}\partial
_{\beta }\xi ^{mn}+\partial _{\alpha }X\partial _{\beta }X+\frac{e^{-2X}}{2}%
\xi ^{mn}\partial _{\alpha }\omega _{m}\partial _{\beta }\omega _{n}.
\end{equation}

\begin{theorem}
The equations 4.14\ are the Einstein equations on $(V_{3},\tilde{g})$ with
source \ a mapping $u:(V_{3},\tilde{g})\rightarrow (R^{D},G)$ by $(x^{\alpha
})\mapsto (X,\xi _{mn},\omega _{m}),$ where $G$ is the riemannian metric: 
\begin{equation}
G\equiv (dX)^{2}+\frac{1}{4}\xi ^{mp}\xi ^{nq}d\xi _{mn}d\xi _{pq}+\frac{1}{2%
}e^{-2X}\xi ^{mn}d\omega _{m}d\omega _{n}.
\end{equation}
\end{theorem}

\begin{proof}
The stress energy tensor of the mapping $u=(X,\xi ,\omega )$ is by
definition 
\begin{equation}
T_{\alpha \beta }=:\partial _{\alpha }u.\partial _{\beta }u-\frac{1}{2}%
\tilde{g}_{\alpha \beta }\partial _{\lambda }u.\tilde{\partial}^{\lambda }u
\end{equation}
where a dot denotes a scalar product in the metric $G.$ The corresponding
right hand side for $\tilde{R}_{\alpha \beta }$ is 
\begin{equation}
\rho _{\alpha \beta }\equiv \partial _{\alpha }u.\partial _{\beta }u
\end{equation}
The identity of this $\rho _{\alpha \beta }$ with 4.15 results from the
definition of $G.$
\end{proof}

\section{Einstein wave map system.}

The Bianchi identities satisfied by the Ricci tensor of $\tilde{g}$ together
with the equations 4.14 imply that the stress energy tensor $T_{\alpha \beta
}$ of the mapping $u$ satisfies the conservation law $\tilde{\nabla}_{\alpha
}\tilde{T}^{\alpha \beta }=0.$ For any map between pseudo riemannian
manifolds it holds that 
\begin{equation}
\tilde{\nabla}_{\alpha }\tilde{T}^{\alpha \beta }\equiv \hat{\nabla}^{\alpha
}\partial _{\alpha }u.\tilde{\partial}^{\beta }u.
\end{equation}
where $\hat{\nabla}^{\alpha }\partial _{\alpha }$ is the wave map operator.
It is straightforward to check that for the considered $\rho _{\alpha \beta
} $, as foreseen due to the consistency of the original vacuum equations,
the vector $\tilde{\nabla}_{\alpha }\tilde{T}^{\alpha \beta }$ takes the
form of the scalar product of $\tilde{\partial}^{\beta }u$ with the
semilinear wave operators found before for $X$, $\xi $ and $\omega .$

We have proved the following theorem, where ''essentially'' means ''with the
choice $H_{m}=0$ of harmonic 1\ forms appearing in 3.2'':

\begin{theorem}
The class of d dimensional vacuum Einstein equations $Ricci(^{(d)}g)=0$ with
a $d-3$ dimensional spacelike \ abelian isometry group are essentially
equivalent to Einstein equations for a lorentzain metric $\tilde{g}$ on $%
V_{3}$ with source a wave map $u:(V_{3},\tilde{g})$ $\rightarrow $ $%
(R^{D},G),$ that is to the system 
\begin{equation}
\tilde{R}_{\alpha \beta }=\partial _{\alpha }u.\partial _{\beta }u,\text{ \
\ \ \ }\hat{\nabla}_{\alpha }\tilde{\partial}^{\alpha }u=0,
\end{equation}
with a dot the scalar product in $G$ and $\hat{\nabla}$ the covariant
derivative$\footnote{{\footnotesize That is }$\hat{\nabla}_{\alpha }\partial
_{\beta }u^{A}\equiv \partial _{\alpha }\partial _{\beta }u^{A}-\tilde{\Gamma%
}_{\alpha \beta }^{\lambda }\partial _{\lambda }u^{A}+\Gamma
_{BC}^{A}\partial _{\alpha }u^{B}\partial _{\beta }u^{C},${\footnotesize \
with }$\tilde{\Gamma}${\footnotesize \ and }$\Gamma ${\footnotesize \ \
connection coefficients respectively of the metrics \~{g}\ and G.}}$
associated with $u.$
\end{theorem}

\section{Einstein Maxwell Higgs equations.}

We return to the case $d=5,$ Einstein-Maxwell-scalar equations, to give the
physical interpretation of the metric $\xi $ wich is now on $R^{2}.$ In the
Kaluza - Klein formulation the spacetime is a Lorentzian manifold $%
(V_{5},^{(5)}g)$ with a $S^{1}$ fiber bundle structure induced by a
spacelike Killing vector field $\partial /\partial x^{4}$ and with base a 4
manifold $V_{4}.$ The metric can be written: 
\begin{equation}
^{(5)}g=:^{(4)}g+e^{2\phi }(dx^{4}+B)^{2}.
\end{equation}
with $^{(4)}g$ a Lorentzian metric$,$ $B$ a locally defined 1 form and $\phi 
$ a scalar field all defined on $V_{4},$ the vacuum Einstein equations for ($%
V_{5},^{(5)}g)$ imply the Einstein Maxwell equations on ($V_{4},^{(4)}g)$
with electromagnetic potential $B$ coupled with the scalar field $\phi .$
Suppose that the gravitational and electromagnetic fields as well as the
additional scalar field are invariant under another 1 - parameter group $%
S^{1}$ endowing $V_{4}$ with a $S^{1}$ fiber bundle structure, with Killing
vector field $\partial /\partial x^{3}$ and base a 3 dimensional manifold $%
V_{3}$ with local coordinates $x^{\alpha },$ $\alpha =0,1,2.$ This
hypothesis is equivalent to the independence of $\phi $ and $B\equiv 
\underline{B}+B_{3}dx^{3}$ on $x^{3},$ with $\underline{B}\equiv B_{\alpha
}dx^{\alpha },$ together with the possibility of writing $^{(4)}g$ under the
following form, all quantities defined on the quotient manifold $V_{3},$%
\begin{equation}
^{(4)}g\equiv \underline{g}+e^{2\gamma }(dx^{3}+A)^{2},\text{ \ }\underline{g%
}\equiv \underline{g}_{\alpha \beta }dx^{\alpha }dx^{\beta },\text{\ \ \ }%
A=:A_{\alpha }dx^{\alpha }.
\end{equation}

The hypotheses are equivalent to saying that the metric $^{(5)}g$ admits a 2
parameter abelian group $G_{2},$ as shown explicitly by the following lemma.

\begin{lemma}
The metric 
\begin{equation}
^{(5)}g=:\underline{g}+e^{2\gamma }(dx^{3}+A)^{2}+e^{2\phi }(dx^{4}+%
\underline{B}+B_{3}dx^{3})^{2}
\end{equation}
reads as a metric on $V_{5}$ with an abelian $G_{2}$ isometry group, namely,
setting $a=:a_{\alpha }dx^{\alpha },$ \ \ \ \ $b=b_{\alpha }dx^{\alpha }:$%
\begin{equation}
^{(5)}g\equiv \underline{g}+\xi _{33}(dx^{3}+a)^{2}+2\xi
_{34}(dx^{3}+a)(dx^{4}+b)+\xi _{44}(dx^{4}+b)^{2}
\end{equation}
with 
\begin{equation}
e^{2\phi }=\xi _{44},\text{ \ }B_{3}=\frac{\xi _{34}}{\xi _{44}},\text{ \ \ }%
e^{2\gamma }=\frac{\det \xi }{\xi _{44}}\equiv \xi _{33}-\frac{\xi _{34})^{2}%
}{\xi _{44}}
\end{equation}
and 
\begin{equation}
A=a,\text{ \ \ }\underline{B}=b+\frac{\xi _{34}}{\xi _{44}}a
\end{equation}
\end{lemma}

\begin{proof}
Straightforward calculation.
\end{proof}

In the case where $B_{3}=0,$ that is when the electromagnetic potential
vector is orthogonal to the Killing vector $\partial /\partial x^{3},$ the
field $\xi $ reduces to: 
\begin{equation}
\xi _{44}=e^{2\phi },\text{ }\xi _{34}=0,\text{ \ \ }\xi _{33}=e^{2\gamma }%
\text{ \ \ }X=\gamma +\phi .
\end{equation}
and the metric $G$ is given by\footnote{{\footnotesize In the case of vacuum
4 dimensional Einstein equations with 1 spacelike isometry group }$\phi
=\omega _{4}=0${\footnotesize \ the metric G reduces to the metric of the
Poincar\'{e} plane used in previous articles.}}: 
\begin{equation}
G\equiv (dX)^{2}+(d\phi )^{2}+(d\gamma )^{2}+\frac{1}{2}e^{-2(X+\phi
)}(d\omega _{4})^{2}+\frac{1}{2}e^{-2(X+\gamma )}(d\omega _{3})^{2}.
\end{equation}

\section{Cauchy problem.}

\subsection{wave map equation}

The wave map equation is a semilinear wave equation on $V_{3}$ when the
metric $\tilde{g}$ is known.

\subsection{Einstein equations.}

Einstein equations on a 3 manifold are essentially\footnote{{\footnotesize %
We will see that they are non dynamical if and only if the surface }$\Sigma $
{\footnotesize is topologically a sphere.}} non dynamical, since the Riemann
tensor is in that case equivalent to the Ricci tensor. To solve the Cauchy
problem for the Einstein equations 5.2 one takes as usual $V_{3}=\Sigma
\times R$ and consider a 2+1 splitting of the metric $\tilde{g}$, with $N$
and $\nu $ its lapse and shift, $g$ a $t-$ dependent riemanian metric on $%
\Sigma :$%
\begin{equation*}
\tilde{g}\equiv -N^{2}dt^{2}+g_{ab}(dx^{a}+\nu ^{a}dt)(dx^{b}+\nu ^{b}dt).
\end{equation*}
We denote by $k$ the extrinsic curvature of a surface $\Sigma _{t}=\Sigma
\times \{t\}$ embedded in $(V_{3},\tilde{g}),$ by $\tau $ its mean extrinsic
curvature. Then, with $h$ a traceless tensor: 
\begin{equation*}
k_{ab}\equiv -\frac{1}{2N}\bar{\partial}_{0}g_{ab}\equiv h_{ab}+\frac{1}{2}%
g_{ab}\tau ,\text{ }\tau =g^{ab}k_{ab}.
\end{equation*}

\subsubsection{Einstein constraints.}

Part of the Einstein equations are the constraints on each $\Sigma _{t},$
that is: 
\begin{equation}
R(g)-k.k+\tau ^{2}=|u^{\prime }|^{2}+g^{ab}\partial _{a}u.\partial _{b}u,%
\text{ \ \ }u^{\prime }=:N^{-1}\partial _{0}u,
\end{equation}
\begin{equation}
\nabla _{b}k_{a}^{b}-\partial _{a}\tau =u^{\prime }.\partial _{a}u.
\end{equation}
To solve these constraints one uses the conformal method, that is we set: 
\begin{equation}
g_{ab}=e^{2\lambda }\sigma _{ab}
\end{equation}
and we take as a gauge condition that $\tau $ is a given function of $t$
alone (CMC gauge).

The constraints become on each $\Sigma _{t},$ with covariant derivative,
laplacian and norms in the metric $\sigma _{t}:$

\begin{itemize}
\item  The equation linear in $h,$ given $\sigma ,$ $u,$ $\dot{u}$%
\begin{equation*}
D_{b}h_{a}^{b}=L_{a}\equiv -D_{a}u.\dot{u},\text{ \ \ }\dot{u}=:e^{2\lambda
}N^{-1}\partial _{0}u
\end{equation*}
One solves this system by setting $h=q+r$ with $h$ a $TT$ (Transverse,
Traceless) tensor, that is a solution of the homogeneous system 
\begin{equation}
D_{b}q_{a}^{b}=0,\text{ \ \ \ }q_{a}^{a}=0,
\end{equation}
and $r$ a conformal Lie derivative of a vector field $Y,$ solution then of
an elliptic system.

\item  The nonlinear elliptic equation for $\lambda $ 
\begin{equation*}
\Delta _{\sigma }\lambda =f(x,\lambda )\equiv \frac{\tau ^{2}}{4}e^{2\lambda
}-\frac{1}{2}p_{2}e^{-2\lambda }+\frac{1}{2}p_{3},
\end{equation*}
\begin{equation*}
p_{2}\equiv \mid \overset{.}{u}\mid ^{2}+\mid h\mid ^{2},\text{ \ \ \ }%
p_{3}\equiv R(\sigma )-|Du|^{2}
\end{equation*}
\end{itemize}

\subsubsection{Equations for lapse and shift.}

The Einstein equation $\tilde{R}_{00}=\rho _{00}$ gives the following
elliptic equation for $N:$%
\begin{equation}
\Delta _{\sigma }N-aN=-e^{2\lambda }\partial _{t}\tau
\end{equation}
\begin{equation}
a\equiv e^{-2\lambda }(\mid h\mid _{g}^{2}+\mid \dot{u}\mid
_{g}^{2})+e^{2\lambda }\frac{\tau ^{2}}{2}
\end{equation}

A linear differential equation for $\nu ,$ depending on $\lambda ,h,N$ and
linearly on $\partial _{t}\sigma ,$ results form the expression of $k.$ This
differential equation admits solutions only if its non homogeneous term is $%
L^{2}$ orthogonal to conformal Killing vector fields.

When $\sigma ,q$ and the mapping $u$ are known on $V_{3}$ the equations for $%
r,\lambda ,N$ are an elliptic system on each $\Sigma _{t}.$

\subsection{Hyperbolic-elliptic system, given $\protect\sigma $ and $q.$}

When the family of metrics $\sigma _{t}$ and TT tensor $q_{t}$ are given,
the wave map $u$ and the metric coefficients $N,$ $\lambda ,$ $\nu $ in CMC
gauge satisfy a coupled hyperbolic - elliptic system with auxiliary unknown $%
r.$ The Bianchi identities show that the space part of $Ricci(\tilde{g}%
)-\rho $ is a TT tensor on $\Sigma _{t}$ when this system is satisfied.

\subsection{Teichmuller parameters ($\Sigma $ compact).}

From now on we consider only the case of a \textbf{compact }$\Sigma .$

When $\Sigma $ is diffeomorphic to $S^{2}$ all the possible metrics $\sigma
_{t}$ are conformal (up to diffeomorphisms) to the canonical metric, and
there does not exist on $(\Sigma ,\sigma )$ any non identically zero TT
tensor hence $q\equiv 0,$ and $Ricci(\tilde{g})-\rho =0$ as soon as
constraints and lapse equations are satisfied. We can take as gauge
condition, in addition to CMC, $\partial _{t}\sigma =0.$ There are
integrability conditions for the shift equation\footnote{{\footnotesize See
CB-M3.}}.

When $\Sigma $ is a 2 torus, i.e. Genus($\Sigma )=0,$ any metric $\sigma $
on $\Sigma $ admits two linearly independent Killing vector fields, and two
linearly independent TT tensors. A metric on $\Sigma $ is conformal to a
metric with zero scalar curvature, one can choose $\sigma $ such that $%
R(\sigma )=0.$ We do not treat this case.

When $G=:$Genus($\Sigma )>1,$ a metric on $\Sigma $ is conformal to a metric
with constant negative scalar curvature, one can choose $\sigma $ such that $%
R(\sigma )=-1.$ The space of \ classes of conformally equivalent metrics on $%
\Sigma ,$ called Teichmuller space $T_{eich},$ can be identified with $%
M_{-1}/D_{0}$, quotient of the space of metrics with $R(\sigma )=-1$ by the
group of diffeomorphisms homotopic to the identity. $M_{-1}\mathcal{%
\rightarrow }T_{eich}$ is a trivial fiber bundle whose base can be endowed
with the structure of the manifold $R^{6G-6}$. As a gauge condition we
impose to the metric $\sigma _{t}$ to be in some chosen cross section $%
Q\rightarrow \psi (Q)$ of this fiber bundle. Let $Q^{I},I=1,...,6G-6$ be
coordinates in $T_{eich}$, then $\partial \psi /\partial Q^{I}$ is a known
tangent vector to $M_{-1}$ at $\psi (Q)$, that is a traceless symmetric
2-tensor field on $\Sigma ,$ sum of a $TT$ tensor field $X_{I}(Q)$ and Lie
derivative of a vector field on $(\Sigma ,\psi (Q))$. Solvability condition
for the shift equation determines $dQ/dt$ in terms of $q_{t}$ and
conversely. One obtains an ordinary differential system for the evolution of 
$Q$ by the $L^{2}$ orthogonality of $\tilde{R}_{ab}-\rho _{ab}$ with the
6G-6 dimensional vector space of $TT$ tensors over $\Sigma .$

\subsection{Conclusion.}

We have to solve the coupled system:

1. Elliptic equations on $(\Sigma ,\sigma _{t})$ with $\sigma _{t}=\psi
(Q(t)),$ with coefficients depending on $u.$

2. ODE on $R$ for $Q(t)$ with coefficients depending on $u$ and elliptic
unknowns.

3. Wave map system on $(\Sigma \times R,\tilde{g})$, $\tilde{g}$ determined
by $\sigma _{t}$ and elliptic unknowns.

\begin{theorem}
(local existence.)The Cauchy data on the compact orientable smooth surface $%
\Sigma _{t_{0}},$ genus($\Sigma )>1$ are:

1. A $C^{\infty }$ metric $\sigma _{0}$ and TT tensor $q_{0}.$

2. Wave map initial data: $u_{0}=u(t_{0},.)\in H_{2},$ $\dot{u}%
_{0}=e^{2\lambda }u^{\prime }(t_{0},.)\in H_{1}.$

The 2+1 Einstein wave map system has a solution taking these Cauchy data and
such that $u\in C^{0}([t_{0},T),H_{2})\cap C^{1}([t_{0},T),H_{1}),$ $\lambda
,N,\nu \in C^{0}([t_{0},T),W_{3}^{p})\cap C^{1}([t_{0},T),W_{2}^{p}),$ 1%
\TEXTsymbol{<}p\TEXTsymbol{<}2, $N$\TEXTsymbol{>}0, if T-t$_{0}$ is small
enough.

There is a corresponding Einteinian \ spacetime $(V_{d},^{(d)}g)$ if the
initial data satisfy the Chern integrability condition for the construction
of $A$ and $B$.
\end{theorem}

\section{Scheme for global existence}

We have chosen to work in CMC gauge: $\tau $ is a time parameter which, in
our conventions taken from MTW inceases from $-\infty $ to $0$ if the
spacetime expands from a big-bang singularity to a moment of maximum
expansion. To have notations more familiar to the analyst we choose as time
parameter $t=-\tau ^{-1},$ then $t$ increases from $t_{0}>0$ to infinity
when $\Sigma _{t}$ expands from $\tau _{0}<0$ to zero. In the case where $%
R(g)<0$ the constraint 7.1 shows that this moment of maximum expansion
cannot be attained. Existence on $(t_{0},\infty )$ will result from a priori
bounds of the norms appearing in local existence theorem. These a priori
bounds result from energy estimates for the wave map and elliptic estimates,
in particular for the confromal factor $\lambda $ which satisfies a non
linear equation for which estimation depends crucially on the negative sign
of $R(\sigma ),$ i.e. by the Gauss Bonnet theorem the fact that Genus($%
\Sigma )>1.$ The estimates involve also the uniform equivalence of $\sigma
_{t}$ with $\sigma _{0}.$ This requires decay in $t$ of the ''total
energy''. This decay is a consequence of the expansion of the metric $g(t,.)$
of $\Sigma _{t},$ obtained when Genus($\Sigma )>1,$ but its proof requires
the introduction of corrected energies, as already in CB-M2. The proofs are
essentially the same as in the vacuum case of CB2, at least for the Einstein
- Maxwell Higgs system with $B_{3}=0$ where the target metric takes a simple
form. We sketch below the main steps that we use.

\section{First energy estimate.}

One defines the first energy not only as the energy of the wave map but by,
with \TEXTsymbol{\vert}.\TEXTsymbol{\vert}$_{g}$ the point wise norm in the
metrics $g$ and $G,$ 
\begin{equation*}
E(t)\equiv \int_{\Sigma _{t}}(|\partial u|_{g}^{2}+|u^{\prime }|^{2}+\frac{1%
}{2}|h|_{g}^{2})\mu _{g}
\end{equation*}
The Hamiltonian constraint 
\begin{equation*}
R(g)+\frac{\tau ^{2}}{2}=|\partial u|_{g}^{2}+|u^{\prime }|^{2}+\frac{1}{2}%
|h|_{g}^{2}
\end{equation*}
together with the Gauss Bonnet formula ($\chi $ the Euler Poincar\'{e}
constant) 
\begin{equation}
\int_{\Sigma _{t}}R(g)\mu _{g}=4\pi \chi
\end{equation}
give, without using wave map equation, that: 
\begin{equation}
\frac{dE(t)}{dt}=\frac{1}{2}\tau \int_{\Sigma _{t}}N(|u^{\prime }|^{2}+\frac{%
1}{2}|h|_{g}^{2})\mu _{g}\leq 0.
\end{equation}
We deduce from this equality that $E(t)$ is a non increasing function of $t$
since $\tau <0,$ but we do not obtain its decay due to the absence of $%
|Du|^{2}$ in the right hand side.

\section{First elliptic estimates.}

The definition of $E(t)$ implies 
\begin{equation*}
||h||_{L^{2}(\sigma )}\leq e^{2\lambda _{M}}E(t)
\end{equation*}
while under hypothesis \ $R(\sigma )=-1$ the maximum principle applied to
... implies\ 
\begin{equation}
e^{2\lambda }\geq 2\tau ^{-2}
\end{equation}
and, with the parameter choice $\tau =-\frac{1}{t},$ it implies, applied to
... \ 
\begin{equation}
\ 0<N\leq 2.
\end{equation}

Further elliptic estimates require bounds of $\partial u.\partial u$ and $%
\partial u.u^{\prime }$ in $W_{1}^{p}(\sigma ),$ $1<p<2,$ which are obtained
in terms of the second energy of the wave map.

\section{Second energy}

We denote by $\hat{\nabla}$ and $\hat{\partial}_{0}$ covariant derivatives
for mappings ($\Sigma ,g)$ or $(R,dt^{2})$ into $(R^{D},G).$ Set: 
\begin{equation}
E^{(1)}(t)\equiv \int_{\Sigma _{t}}(\mid \hat{\Delta}_{g}u\mid ^{2}+\mid 
\hat{\nabla}u^{\prime }\mid ^{2})\mu _{g}
\end{equation}
\begin{equation}
\varepsilon ^{2}=:E(t),\text{ \ \ }\varepsilon _{1}^{2}=:\tau ^{-2}E_{1}(t).
\end{equation}
One finds after long computations using elliptic estimates applied to the
constraints and lapse equations that: 
\begin{equation}
1\leq \frac{1}{\sqrt{2}}|\tau |e^{\lambda }\leq 1+C_{E,\sigma }(\varepsilon
+\varepsilon _{1}),
\end{equation}
\begin{equation}
0\leq 2-N\leq C_{E,\sigma }(\varepsilon ^{2}+\varepsilon \varepsilon _{1})
\end{equation}
where we denote by $C_{\sigma ,E}$ numbers depending only on a priori bound
of $\varepsilon $ and $\varepsilon _{1},$ and on the domain of $\sigma $ in $%
Teich$ supposed to be compact.

Using these bounds and the wave map $(\Sigma \times R,\tilde{g})\rightarrow
(R^{D},G)$ which reads in our notations: 
\begin{equation}
-N^{-1}\hat{\partial}_{0}\partial _{0}u+g^{ab}\hat{\nabla}_{a}(N\partial
_{b}u)+Nu^{\prime }=0.
\end{equation}
we find that: 
\begin{equation}
\frac{dE^{(1)}}{dt}-2\tau E^{(1)}=\tau \int_{\Sigma _{t}}N\mid Du^{\prime
}\mid ^{2}\mu _{g}+Z\leq Z
\end{equation}
with: 
\begin{equation}
|Z|\leq C_{\sigma ,E}(\varepsilon +\varepsilon _{1})^{3}.
\end{equation}
\begin{equation}
|Z|\leq |\tau |^{3}C_{\sigma ,E}(\varepsilon +\varepsilon _{1})^{3}.
\end{equation}
The inequalities 11.6 and 9.2, are not sufficient to prove the bound of $%
\varepsilon +\varepsilon _{1}$ and the fact that $\sigma _{t}$ projects on a
fixed compact subset of $T_{eich},$ the proof of this last property requires
decay of $\varepsilon +\varepsilon _{1}$.

\section{Corrected energies.}

To obtain the decay property one introduces corrected energies and exploit
the negative (non definite) terms in the energies inequalities:

\begin{equation}
E_{\alpha }(t)=E(t)-\alpha \tau \int_{\Sigma _{t}}(u-\overset{\_}{u}%
).u^{\prime }\mu _{g}
\end{equation}
\begin{equation*}
\overset{\_}{u}=\frac{1}{Vol_{\sigma _{t}}}\int_{\Sigma _{t}}u\mu _{\sigma },
\end{equation*}
\begin{equation}
E_{\alpha }^{(1)}(t)=E^{(1)}(t)+\alpha \tau \int_{\Sigma _{t}}\hat{\Delta}%
_{g}u.u^{\prime }\mu _{g}
\end{equation}

The use of elliptic estimates leads to 
\begin{equation}
\frac{dE_{\alpha }}{dt}-k\tau E_{\alpha }\leq |\tau |C_{\sigma
,E}(\varepsilon +\varepsilon _{1})^{3},
\end{equation}
\begin{equation}
\frac{dE_{\alpha }^{(1)}}{dt}-(2+k)\tau E_{\alpha }^{(1)}+|\tau
|^{3}C_{\sigma ,E}(\varepsilon +\varepsilon _{1})^{3}.
\end{equation}
We denote by $\Lambda _{\sigma }$ the first positive eigenvalue of $-\Delta
_{\sigma }$ and we prove that $E_{\alpha }+\tau ^{-2}E_{\alpha }^{(1)}$ is
equivalent to the total energy $\varepsilon ^{2}+\varepsilon _{1}^{2}$ under
the following conditions:

\begin{itemize}
\item  
\begin{equation}
\alpha =\frac{1}{4},\text{ \ }k=1\text{ \ if \ }\Lambda _{\sigma }>\frac{1}{8%
}
\end{equation}

\item  
\begin{equation*}
\alpha <\frac{4}{8+\Lambda _{\sigma }^{-1}}\text{\ , \ }0<k<1,\text{ \ if }%
\Lambda _{\sigma }\leq \frac{1}{8}.
\end{equation*}
\end{itemize}

All these differential inequalities imply the inequalities: 
\begin{equation}
(\varepsilon ^{2}+\varepsilon _{1}^{2})(t)\leq t^{-k}M_{1}(\varepsilon
^{2}+\varepsilon _{1}^{2})(t_{0}).
\end{equation}

\section{Future complete existence (non linear stability)}

\begin{theorem}
Let ($\sigma _{0},q_{0})\in C^{\infty }(\Sigma _{0})$ and ($u_{0},\overset{.%
}{u}_{0})$ $\in H_{2}(\Sigma _{0},\sigma _{0})$ $\times H_{1}(\Sigma
_{0},\sigma _{0})$ be initial data on the compact manifold $\Sigma _{0},$
Genus($\Sigma _{0})>1,$ satisfying the Chern integrability condition. There
exists a number $\eta >0$ such that, if $E_{tot}(t_{0})<\eta $, the 4
dimensional Einstein Maxwell Higgs system with $S^{1}$ isometry group and
electomagnetic field orthogonal to the Killing field have a solution on $%
\Sigma \times S^{1}\times \lbrack t_{0},\infty ),$ $t=-\tau ^{-1},$ with
initial values determined by $\sigma _{0},q_{0},u_{0},\overset{.}{u}_{0}.$
This space time is globally hyperbolic, future timelike and null complete.
\end{theorem}

\begin{proof}
Using the differential equation satisfied by $Q$ and the decay of the total
energy proved using its a priori bound one obtains the inequality: 
\begin{equation}
|Q(t)-Q(t_{0})|\leq M_{2}(\varepsilon ^{2}+\varepsilon _{1}^{2})(t_{0}).
\end{equation}
This inequality together with 12.6 give a bound of the total energy and of $Q
$ by a bootstrap argument if the initial total energy small enough. One
deduces the existence of the solution on $\Sigma \times \lbrack t_{0},\infty
),$ and the existence for an infinite proper time along the lines \{$%
x\}\times R$ after estimating the usual $H_{2}$ norms in terms of the
geometrically defined second energy. This estimate depends, as in the proof
given in CB2, on the fact that the Riemann curvature of the target metric is
negative. The special form of this metric plays also a role in the estimate
of the second corrected energy.

The global hyperbolicity and completeness is a particular case of a theorem
proved in CB-Cotsakis.
\end{proof}

We have not checked the corresponding properties for the general $G$ given
by 4.16, but we conjecture that the proof goes through in that general case.
We thank V.\ Moncrief and T. Damour for pointing out in discussions at the
Institut des Hautes Etudes Scientifiques in Bures sur Yvette that the metric 
$\xi ^{mp}\xi ^{nq}d\xi _{mn}d\xi _{pq}$ on $R^{3}$ is the product of a line
by the Poincar\'{e} plane.

\textbf{References}

[A-M-T] L.\ Andersson, V.\ Moncrief and A. Tromba On the global evolution
problem in 2+1 gravity J.\ Geom. Phys. 23 1997 n$%
{{}^\circ}%
3-4$,1991-205

[CB1] Y.\ Choquet-Bruhat Global wave maps on curved spacetimes, in
''Mathematical and Quantum Aspects of Relativity and Cosmology'' Cotsakis
and Gibbons ed. LNP 535, Springer 1998, 1-30

[CB2] Y.\ Choquet-Bruhat Future complete Einsteinian spacetimes with U(1)
isometry group, the unpolarized case, in The Einstein equations and the
large scale behaviour of gravitational fields, P. Chrusciel and H.\
Friedrich ed. 2004 Birkhauser 251-298.

[CB-Co] Y.\ Choquet-Bruhat and S. Cotsakis Global hyperbolicity and
completeness, J.\ Geom. Phys. 43 $n%
{{}^\circ}%
4$, 2002, 345-350.

[CB-DM] Y.\ Choquet-Bruhat and C DeWitt-Morette Analysis Manifolds and
Physics II enlarged edition (2000).

[CB-M 1]\textbf{\ }Y.\ Choquet-Bruhat and V.\ Moncrief, Future global in
time einsteinian spacetimes with U(1) isometry group Ann. Henri.
Poincar\'{e} 2 (2001), 1007-1064.

[CB-M 2]\textbf{\ }Y.\ Choquet-Bruhat and V.\ Moncrief, Non linear stability
of einsteinian spacetimes with U(1) isometry group, in Partial differential
equations and mathematical physics, in honor of J. Leray, Kajitani and
Vaillant ed. Birkh\"{a}user.

[CB-M 3] Y.\ Choquet-Bruhat and V.\ Moncrief Existence theorem for solutions
of Einstein equations with 1 parameter spacelike isometry group, Proc.
Symposia in Pure Math, 59, 1996, H.\ Brezis and I.E.\ Segal ed. 67-80.

[CB-Yo] Y.\ Choquet-Bruhat and J.\ W.\ York, Constraints and Evolution in
cosmology, in Cosmological crossroads, 2001, S. Cotsakis and E.
Papantonopoulos ed. 29-58

[L] A. Lichn\'{e}rowicz, Les th\'{e}ories de la gravitation et de
l'\'{e}lectromagn\'{e}tisme, Masson 1955.

[Ma] P.\ O.\ Mazur A relationship betaween the electrovacuum Ernst equation
and non linear $\sigma -$ model Acta Phys. Pol. B14, 1983, 219-234.

[M1] V.\ Moncrief Reduction of Einstein equations for vacuum spacetimes with
U(1) spacelike isometry group, Annals of Physics 167 (1986), 118-142

[M2] V. Moncrief Reduction of the Einstein - Maxwell and the Einstein -
Maxwell - Higgs equations for cosmological spacetimes with U(1) spacelike
isometry group. Class. Quantum Grav. 7 (1990) 329-352.

{\footnotesize Academie des Sciences, 23 quai Conti 75006 Paris Email
YCB@CCR.jussieu.fr}

\end{document}